\documentclass{article}

\usepackage{graphicx}
\usepackage{fancyhdr}
\begin{document}

\title{Finite temperature $SU(2)$ gauge theory: critical coupling and universality class}

\author{Alexander Velytsky\\
\emph{\small
Enrico Fermi Institute,}\\
\emph{\small University of Chicago, 5640 S. Ellis Ave., Chicago, IL 60637, USA}\\
\emph{\small and}\\
\emph{\small
HEP Division and Physics Division,}\\
\emph{\small
Argonne National Laboratory, 9700 Cass Ave., Argonne, IL 60439, USA}
}

\maketitle
\thispagestyle{fancy}

\begin{abstract}
We examine $SU(2)$ gauge theory in $3+1$ dimensions at finite temperature in the
vicinity of critical point. For various lattice sizes in time 
direction ($N_\tau=1,2,4,8$) we extract high precision values of the inverse critical coupling and 
critical values of the 4-th order cumulant of Polyakov loops (Binder cumulant). We check the universality class of the theory by comparing the cumulant values to that of the $3D$ Ising model and find very good agreement.

The Polyakov loop correlators for the indicated lattices are also measured and the string tension values extracted. The high precision values of critical coupling and string tension allow us to study the scaling of dimensionless $T_c/\sqrt{\sigma}$ ratio. The violation of scaling by $<10\%$ is observed as the coupling is varied from weak to strong coupling regime.

\end{abstract}

\section{Introduction}
The original motivation for this study was to re-derive the values of critical coupling for finite temperature
$SU(2)$ gauge theory formulated on lattices with different time-like extent $N_\tau$ at the highest precision possible with modern computational resources (we utilize a small 20-30 node computer cluster). 
For this we rely on the property of universality of $SU(2)$ gauge theory 
at the second order critical point. We adopt a standard for such models procedure of locating the 
position of critical point, which amounts to measuring the 4th order (Binder) cumulant\cite{Binder:1981} $g_4$ for Polyakov loops $P$
\begin{equation}
g_4=1-\frac{\langle P^4\rangle}{3\langle P^2\rangle^2}, \quad P=\frac1{N_\sigma^3}\sum_{\vec{x}}
\frac12{\rm Tr}\prod_{\tau=1}^{N_\tau}U_{\tau,\vec{x};0}
\end{equation}
on $N_\tau\times N_\sigma^3$ lattices in the vicinity of phase transition. 
Note that we use the original form of the Binder cumulant
which differs by a constant factor $1/3$ from the normalized version frequently used in lattice gauge theory literature. 
The finite size scaling (FSS)
of $g_4$ with lattice size $N_\sigma$ in the vicinity of the critical point is known\cite{Binder:1981sa,Binder:2001ha}

\begin{equation}
g_{4,N_\sigma}\approx g_{4,\infty} (1+a_1t N_\sigma^{1/\nu}+a_2N_\sigma^{-y_1}+\cdots)
\label{eq:gscale}
\end{equation}
where $t=(T-T_{c,\infty})/T_{c,\infty}$ is the reduced temperature and $y_1\equiv -\omega>0$ is the exponent of the largest irrelevant scaling field.

The scaling of the temperature value of the intersection point of Binder cumulant curves 
$g_4(t)$ on $N_\sigma$ and $N^{\prime}_\sigma=bN_\sigma, \, b>1$ lattices is
\begin{equation}
t^{*}=-\frac{a_2}{a_1}N_\sigma^{-y_1-1/\nu}\frac{1-b^{-y_1}}{1-b^{1/\nu}}.
\label{eq:tscale}
\end{equation}
It is convenient to define $L_b=N_\sigma^{-y_1-1/\nu}(1-b^{-y_1})/(1-b^{1/\nu})$, so
that (\ref{eq:tscale}) becomes a simple linear function of $L_b$.
Substituting $t^*$ into (\ref{eq:gscale}) results in the scaling for the intersection $g_4$ value
\begin{equation}
g^*_4=g_{4,\infty} \left(1+a_2N_\sigma^{-y_1}\frac{1-b^{-y_1-1/\nu}}{1-b^{-1/\nu}}\right).
\label{eq:gscale2}
\end{equation}
Here we define $L_g=N_\sigma^{-y_1}(1-b^{-y_1-1/\nu})/(1-b^{-1/\nu})$, so that
(\ref{eq:gscale2}) is linear in $L_g$.

The well-known renormalization group relationship allows us to connect the lattice spacing to the coupling $\beta=2N_c/g^2$, ($N_c=2$)
\begin{equation}
a\Lambda_{N_\sigma}=\left(\frac\beta{2N_cb_0}\right)^{b_1/2b_0^2}\exp\left(-\frac\beta{4N_cb_0}\right).
\label{eq:a_scale}
\end{equation}
Using the fact that  the temperature is $T=1/(N_\tau a)$ one obtains the following expression for the reduced temperature in the vicinity of the transition point
\begin{equation}
t=\frac{\beta-\beta_{c,\infty}}{4N_cb_0}\left(1-\frac{2N_cb_1}{b_0}\beta^{-1}_{c,\infty}\right)+
O((\beta-\beta_{c,\infty})^2).
\end{equation}
In the studies of $N_\tau=8$ and 16 phase transitions by Fingberg et al.\cite{Fingberg:1992ri}
no deviations from the leading term were found. Therefore one can assume linear correspondence
between the reduced temperature and reduced inverse coupling and perform the scaling studies
in terms of the inverse lattice coupling $\beta$.

In their renowned work Svetitsky and Yaffe\cite{Svetitsky:1982gs} conjectured that the universality class
of the $d+1$-dimensional $SU(2)$ gauge theory is the $d$-dimensional Ising model. Since then it was confirmed in numerous numerical simulations, e.g. Ref. \cite{Engels:1989fz} for $N_\tau=4$ theory.
Indeed, we observe that for all lattices in the thermodynamic limit the $g_4$ curves intersect in the vicinity of the $3D$ Ising value, albeit with different degree of accuracy. 

In section \ref{sec:mc} we measure values of Binder cumulant in the vicinity of the critical point and determine critical values of inverse coupling and Binder cumulant for $N_\tau=1,2,4,8$ lattices. Our treatment of  different $N_\tau$ lattices is not uniform. We conduct the FSS study for $N_\tau=4$ and 2 lattices, while relatively large $N_\sigma$ values used for $N_\tau=1$ studies prevented us from observing a significant scaling behavior. The $N_\tau=8$ lattice allows for the FSS study, however it is very expensive to simulate and therefore is studied assuming the Ising universality class.

The small $N_\tau=1$ and 2 lattices are of special interest in decimation studies, since the iterative block-spinning procedure has to be stopped when the smallest (or next to smallest) lattice is reached.
Therefore for $N_\tau=1$ lattice we consider a specific lattice formulation suitable for decimation.

In section \ref{sec:string} we  study the quark-antiquark static potential and extract the string tension for $N_\tau/3\ge2$ lattices at temperature $T=T_c/3$. The  knowledge of critical couplings allows us to construct the dimensionless ratio $T_c/\sqrt{\sigma}$ and study its scaling with coupling. We observe relatively small $<10\%$ violation of scaling at strong coupling $\beta=1.8738$ value.

\section{Finite temperature phase transition: Monte Carlo study\label{sec:mc}}
In all simulations performed in this study we use the standard Wilson action. Per one updating sweep
we perform two overrelaxation steps and one heatbath update. 
We use a standard acceptance improved heat bath updating procedure\cite{Fabricius:1984wp,Kennedy:1985nu}. 
Measurements are performed every $2-20$ sweep. We use $10-80$ independent runs (at different initial random generator seeds), each run is averaged into a single bin. This allows us to gain better statistics and avoid autocorrelations. All errors
are computed with the jack-knife method with respect to these bins, except when indicated differently. For various coupling values
$\beta=4/g^2$ in the visinity of the finite temperature confinement-deconfinement phase 
transition after initial equilibration we measure Polyakov loops and compute the Binder cumulant.
The number of sweeps needed for the system to reach equilibrium is estimated by observing the 
Monte Carlo time evolution of the Polyakov loop and plaquette estimates for each of the lattice sizes and typical equilibration times are $\ge10^3$ sweeps.

\subsection{$N_\tau=4$ lattice}
We start with perhaps the most studied lattice $N_\tau=4$. At this point we do not need to assume any particular universality class. We simulate $N_\sigma=8,10,16,20,24$ and $32$ lattices, with  
typical statistics of $30\times 40000$ configurations.

\begin{figure}[ht]
\includegraphics[width=\columnwidth]{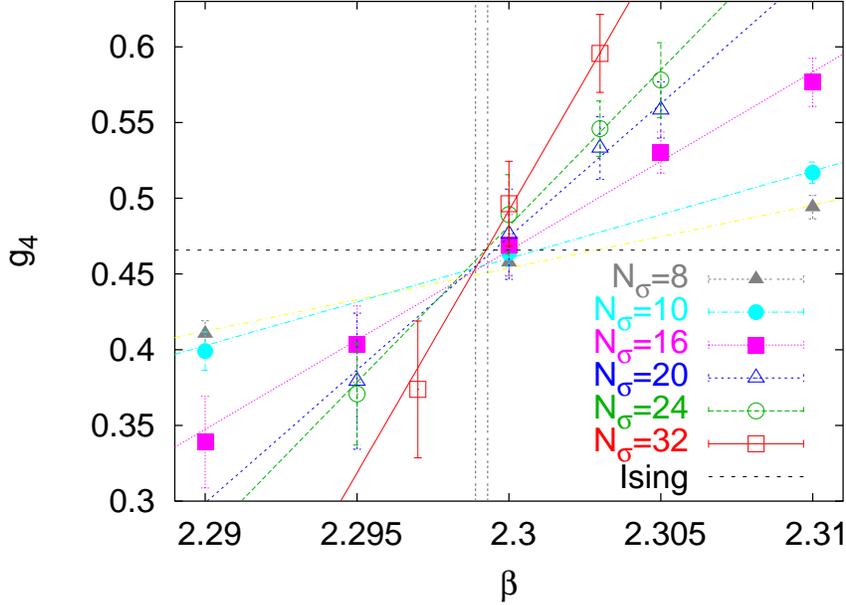}
\caption{The Binder cumulant $g_4$ for $N_\tau=4$, $N_\sigma=8,10,16,24,32$ lattices. Linear fits are represented by solid lines, 
the $T_c$ estimate with errors by vertical lines.}
\label{fig:BinderC4}
\end{figure}

\begin{table}[h]
\begin{center}
\begin{tabular}{|c|cccccc|}
$N_\sigma$&8&10&16&20&24&32\\\hline
$a$&4.17(29)&5.77(36)  &11.80(56)  &17.6(1.2) &20.6(1.3) &34.7(1.9)\\
$b$&-9.13(66)&-12.89(82) &-26.7(1.3) &-39.9(2.8)&-46.9(2.0)&-79.3(4.3)\\
\end{tabular}
\end{center}
\caption{Resulting parameters of the $\chi^2$ fit of the Binder cumulant to a linear  $g_4(\beta)=a\beta+b$ function.}
\label{tab:fit_par4}
\end{table}
\begin{figure}[ht]
\includegraphics[width=\columnwidth]{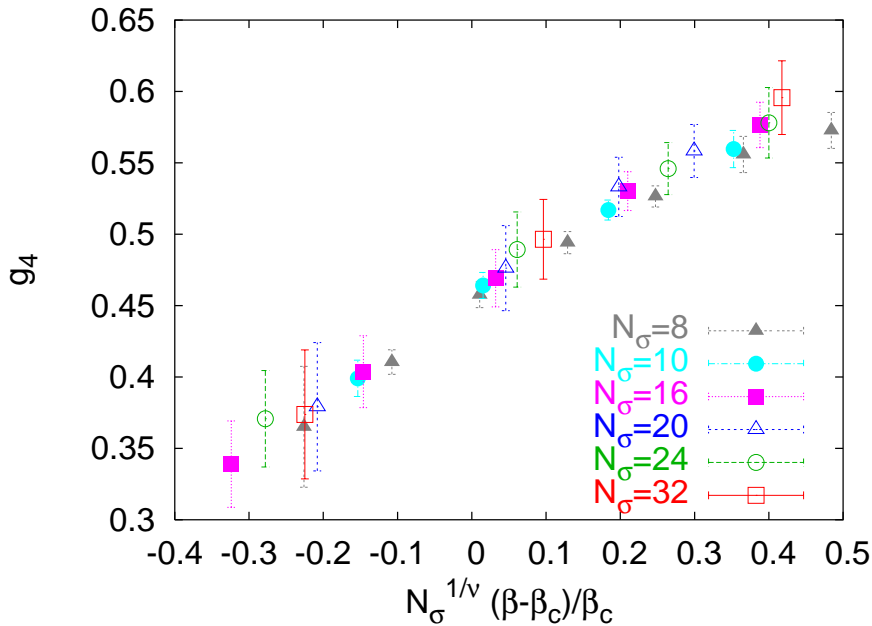}
\caption{The rescaled Binder cumulant $g_4(N_\sigma^{1/\nu}(\beta-\beta_c)/\beta_c)$ for $N_\tau=4$, $N_\sigma=8,10,16,24,32$ lattices.}
\label{fig:BinderC4sc}
\end{figure}
The results for the Binder cumulant measurement are presented in Fig. \ref{fig:BinderC4}  together 
with the fitting lines $g_4(\beta)=a\beta+b$ (see Tab. \ref{tab:fit_par4} for fitting parameters).
\begin{table}[h]
\begin{center}
\begin{tabular}{c|ccccc}
$N_\sigma$ &8              &10            &16               &20&24\\\hline
10&2.2960(19)&*               &*                  &*&*\\
16&2.2985(40)&2.2992(5)&*                  &*&*\\
20&2.2985(44)&2.2988(5)&2.2984(12)&*&*\\
24&2.2983(28)&2.2986(3)&2.2981(7) &2.2977(29)&*\\
32&2.2987(11)&2.2989(1)&2.2988(2)&2.2990(4)&2.2993(3)\\
\hline\hline
10&0.4370(87)&*&*&*&*\\
16&0.4476(19)&0.4556(41)&*&*&*\\
20&0.4474(20)&0.4533(38)&0.446(15)&*&*\\
24&0.4468(14)&0.4520(31)&0.4433(99)&0.433(56)&*\\
32&0.4486(08)&0.4540(22)&0.4514(46)&0.457(11)&0.466(9)\\
\end{tabular}
\end{center}
\caption{\label{tab:cross4} Pairwise intersection coordinates in $g_4-\beta$ plane: the inverse lattice coupling (top) and the 4th cumulant (bottom); $N_\tau=4$.}
\end{table}
The data is fitted to a straight line\footnote{
We have found that the reweighting of Polyakov loops to new $\beta$ values has extremely short range. Therefore we
do not use reweighting and rather use the linear fitting. }  in the vicinity of the 
transition point with a standard $\chi^2$ procedure. The errors in the fitting parameters $a$ and $b$ are
obtained by error propagation.
The resulting goodness of fit in all cases is $Q\approx 0.6-0.9$, which confirms that at the considered 
$\beta$ values the $g_4$ curves are well approximated by lines.
In Tab. \ref{tab:cross4} we list the values $\beta^*$ and $g_4^*$ of intersection points for different 
pairs of lines ($N_\sigma$, $N^{\prime}_\sigma=bN_\sigma$). 

First we check the assumed scaling behavior by plotting the rescaled Binder cumulant 
$g_4(N_\sigma^{1/\nu}(\beta-\beta_c)/\beta_c)$ for various lattice sizes, see Fig. \ref{fig:BinderC4sc}.
For the critical coupling value we use $\beta_c=2.2991$, which we will obtain later in the FSS study.
Indeed, we observe that the curves fall on top of each other. The only noticeable deviation from 
the scaling behavior can be observed for the smallest considered lattice $N_\sigma=8$ far away from the transition point. This is obviously due to the effects from next to leading order scaling terms.
The important observation is that in the region where we perform linear fits there is no deviation from
the scaling for all the lattices and also apparently the linearity holds.

\begin{figure}[ht]
\includegraphics[width=\columnwidth]{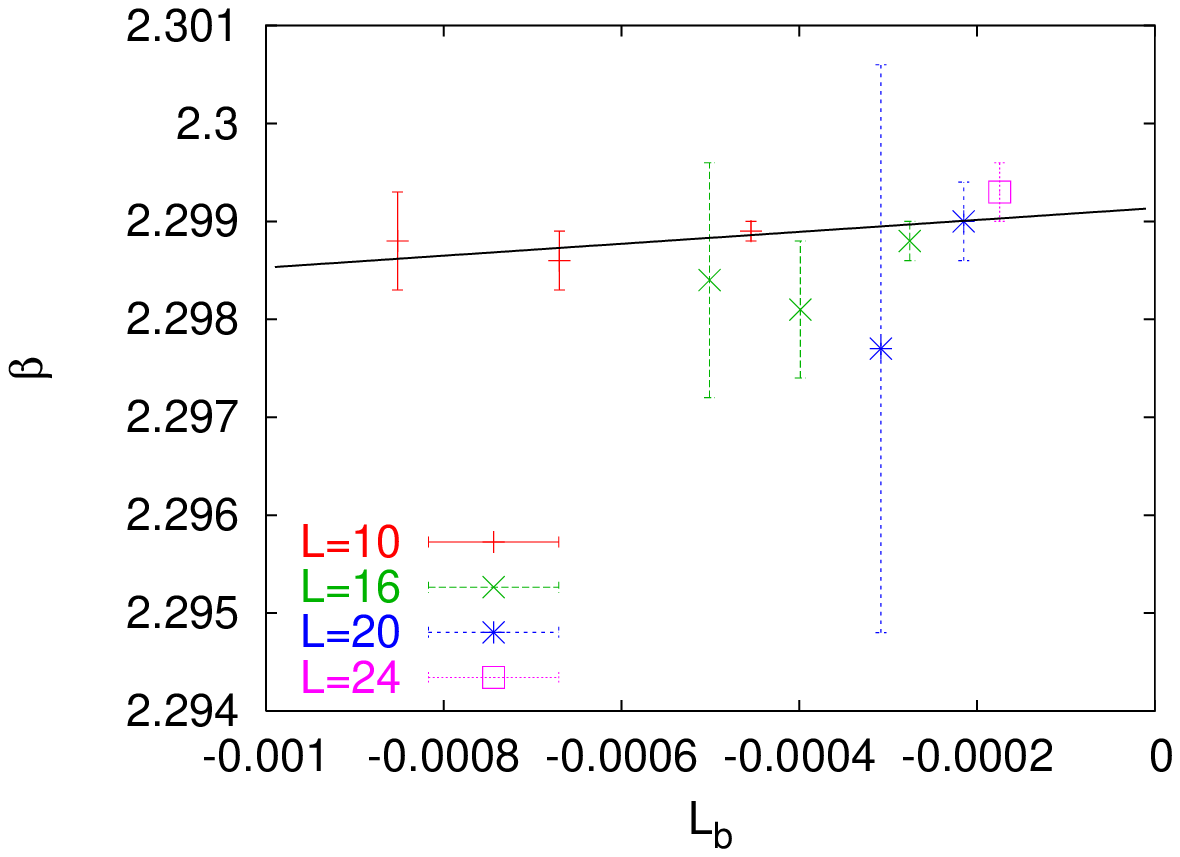}
\caption{The $\beta^*$ coordinates of the pair-wise intersection points for $N_\sigma\equiv L=10,16,20,24$ and various $b$ lattices. The solid line represents a linear fit for all the presented data.  $N_\tau=4$. }
\label{fig:nt4bc}
\end{figure}

\begin{figure}[ht]
\includegraphics[width=\columnwidth]{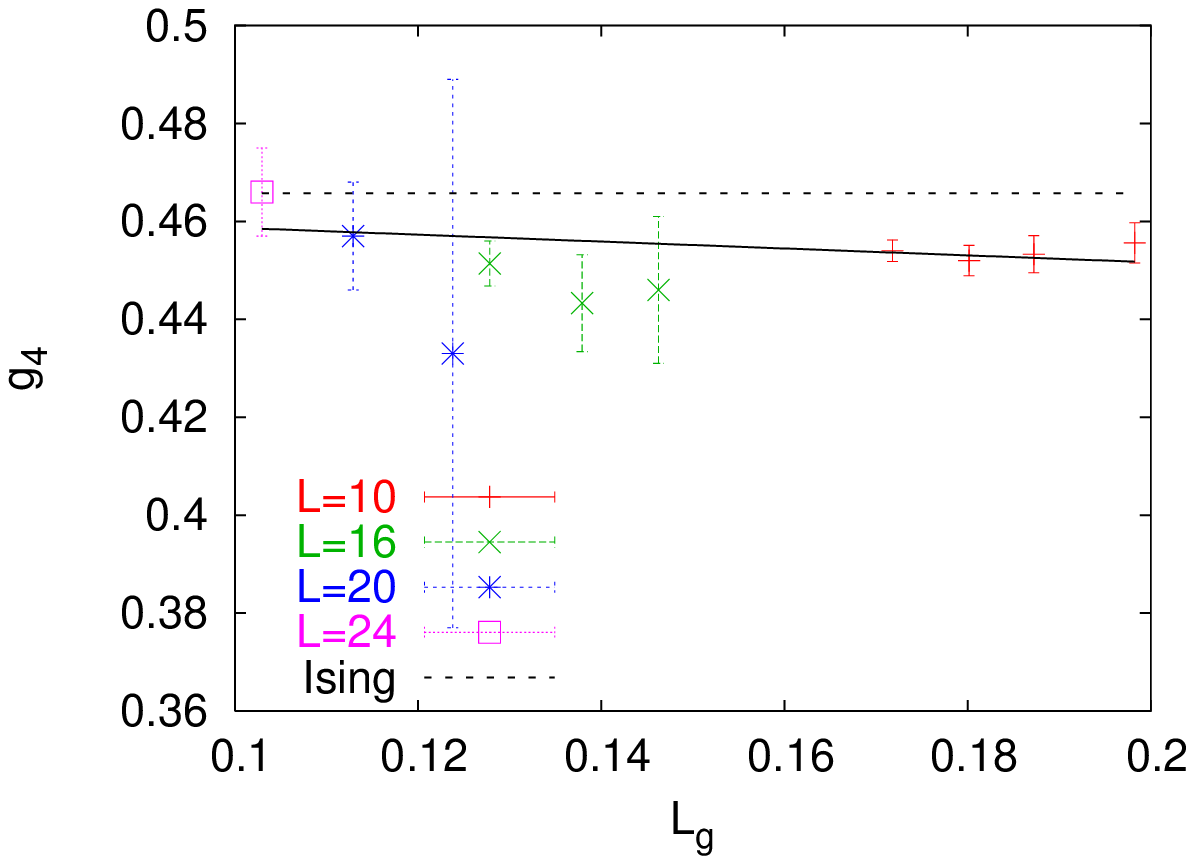}
\caption{Same as Fig. \ref{fig:nt4bc}, but for the Binder cumulant. The horizontal line is the Ising 
$g_4^{Ising}=0.46575$ value.}
\label{fig:nt4g4}
\end{figure}

Next we study the scaling
of the pair-wise intersection point coordinates according to (\ref{eq:tscale}) and (\ref{eq:gscale2}).
First we look at the $\beta^*$ coordinate of intersection points versus $L_b$, see Fig. \ref{fig:nt4bc}, 
and then at the $g^*_4$ coordinte of intersection points versus $L_g$, see Fig. \ref{fig:nt4g4}.
For this we need to know the values of critical exponents $\nu$ and $y_1$. As we will see 
(from the quality of data) it is not practical to extract them from the data, instead we take them to be equal to the Ising values known 
with good accuracy $1/\nu=1.5887(85)$ and $y_1=0.812$ \cite{Ferrenberg:1991}. For a self-consistency check we verify that the scaling resulting from use of these values is adequate.

For each smaller lattice size from $N_\sigma=10,16,20,24$ set (represented by
a different point type/color on the figure) we take various possible $b$ values, which index the larger
lattice.
As one can see from Fig. \ref{fig:nt4bc} despite the fact that the simulation statistics is sufficient for the precise location
of the intersection point the error bars can become large as intersecting lines approach the collinear limit. We performed 
a linear fit in accordance with (\ref{eq:tscale}). The goodness of fit is $Q=0.66$ which suggests that
the scaling equation is correct. The limit $L_b\rightarrow 0$ corresponds to the thermodynamic limit
and yields $\beta_{c,\infty}=2.2991(2)$. Note that the largest
lattices intersection ($N_\tau=24$ and $N_\tau^\prime=32$) happens at $\beta^*=2.2993(3)$ which agrees with the thermodynamic limit result. These results should be 
compared with the value $\beta_c=2.2985(6)$ of Ref. \cite{Engels:1989fz} or $2.2986(6)$ (intersection 12 and 18) of Ref.
\cite{Engels:1992fs} and indeed we find a very 
good agreement.

The value of Binder cumulant for the 3D Ising model is well known. In the lattice gauge theory literature the early estimate of Ferrenberg et al. $g_4^{Ising}=0.470(5)$\footnote{The error is from the published figure.} \cite{Ferrenberg:1991} is often used. In this work we are using the more recent and more accurate estimate of Hasenbusch et al. \cite{Hasenbusch:1999prb}  $Q\equiv<m^2>^2/<m^4>=
0.62393(13)$ (here $m$ is the Ising model magnetization), which translates into 
$g_4^{Ising}=0.46575(11)$.

The data of Tab. \ref{tab:cross4} and Fig. \ref{fig:BinderC4} strongly support the fact that 
the Binder cumulant reaches the Ising value for the larger lattice intersections. Therefore it is safe
to assume that within the accuracy of this study the $N_\tau=4$ model is indeed in the 3D Ising model universality class.

We continue the study of the scaling of pair-wise intersection point coordinates with the size of participating lattices. Now similarly to $\beta^*$ analysis we plot the $g_4^*$ values of intersection points versus $L_g$, see Fig. \ref{fig:nt4g4}.  As one can see the data does not allow to perform a quality
fit especially in the thermodynamic limit $L_g\rightarrow 0$. Since we already assumed that the Ising universality class holds we may fit $g_4$ directly to $a*\beta+g_4^{Ising}$. The reduction of the
degrees of freedom allows for a better defined fit. The goodness of fit is $Q=0.80$ meaning
that the scaling function is plausible. Also it represents a self-consistency check for the assumption of the Ising universality class.

At this point it is worth to mention that in the studies of the Dyson's hierarchical model \cite{yannick_priv}
it was observed that a too coarse scale in the $\beta$ sampling may lead to inaccurate locations of the intersections when linear fits are used. This inaccuracy may generate sizable errors in the $L_b\rightarrow 0$ and $L_g\rightarrow 0$ limits.

As a side note we would like to point that insufficient statistics prevented us from performing the fits inside separate bins and thus obtaining the relevant errors by use of the jack-knife method on this bins.
Instead the fits were performed on a whole set and the errors were obtained by simple 
error propagation, which may undermine their accuracy.

\subsection{$N_\tau=2$ lattice}
Next we study $N_t=2$ lattice. This lattice is somewhat special because unlike larger $N_{\tau}$
lattices as a result of periodic boundary conditions in $T$ direction every pair of spatial links from different time slices is connected only by a single time-like link.
\begin{table}[h]
\begin{center}
\begin{tabular}{c|ccccc}
$N_\sigma$ &10            &16               &24\\\hline
16&1.87331(2)&*&*\\
24&1.87338(1)&1.87343(2)&*\\
32&1.873422(5)&1.87345(1)&1.87348(2)\\
\hline\hline
16&0.4558(3)&*&*\\
24&0.4565(2)&0.4578(4)&*\\
32&0.4568(1)&0.4583(2)&0.4595(10)\\
\end{tabular}
\end{center}
\caption{\label{tab:cross2} Pairwise intersection coordinates in $g_4-\beta$ plane: the inverse lattice coupling (top) and the 4th cumulant (bottom); $N_\tau=2$.}
\end{table}

\begin{figure}[ht]
\includegraphics[width=\columnwidth]{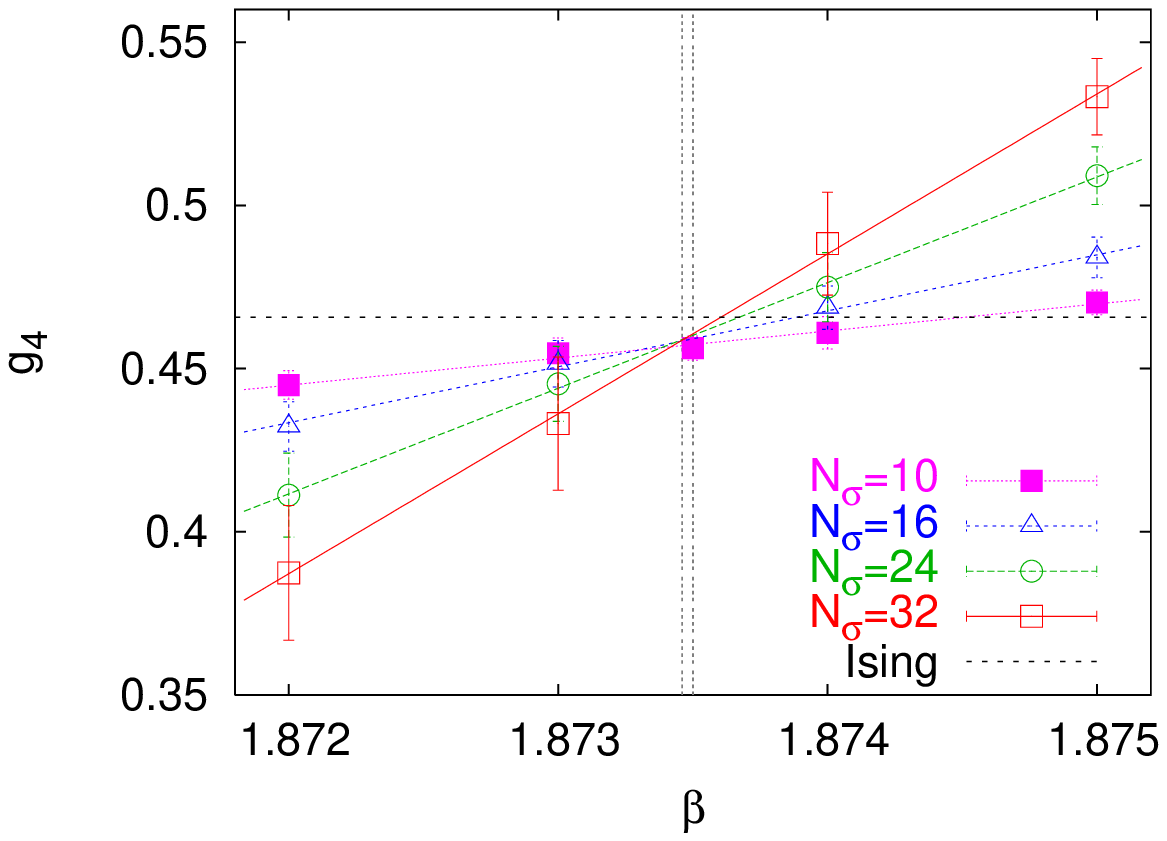}
\caption{The Binder cumulant $g_4$ for $N_\tau=2$, $N_\sigma=10,16,24,32$ lattices. Linear fits are represented by solid lines, 
the $T_c$ estimate with errors by vertical lines.}
\label{fig:BinderC2}
\end{figure}
For this lattice we generated on average from $10\times 0.5*10^5$ configurations for $N_\sigma=16$ to $10\times0.3*10^5$ configurations for $N_\sigma=32$ lattices. The results for Binder cumulant measurements in the vicinity of transition point are presented in Fig. \ref{fig:BinderC2}. As one can see
from the figure the linear fit lines intersect clearly below the Ising value and at what appears to be a single point. We collect the coordinates of the intersection points of the fitting lines in Tab. \ref{tab:cross2}.
We observe that for the intersection coordinate $\beta^*$ of larger lattices there is very insignificant scaling change. Therefore for this quantity we do not perform the FSS analysis and take as the transition 
coupling the value of intersection of lines of the largest 32 and 24 lattices: $\beta_c=1.87348(2)$.

\begin{figure}[ht]
\includegraphics[width=\columnwidth]{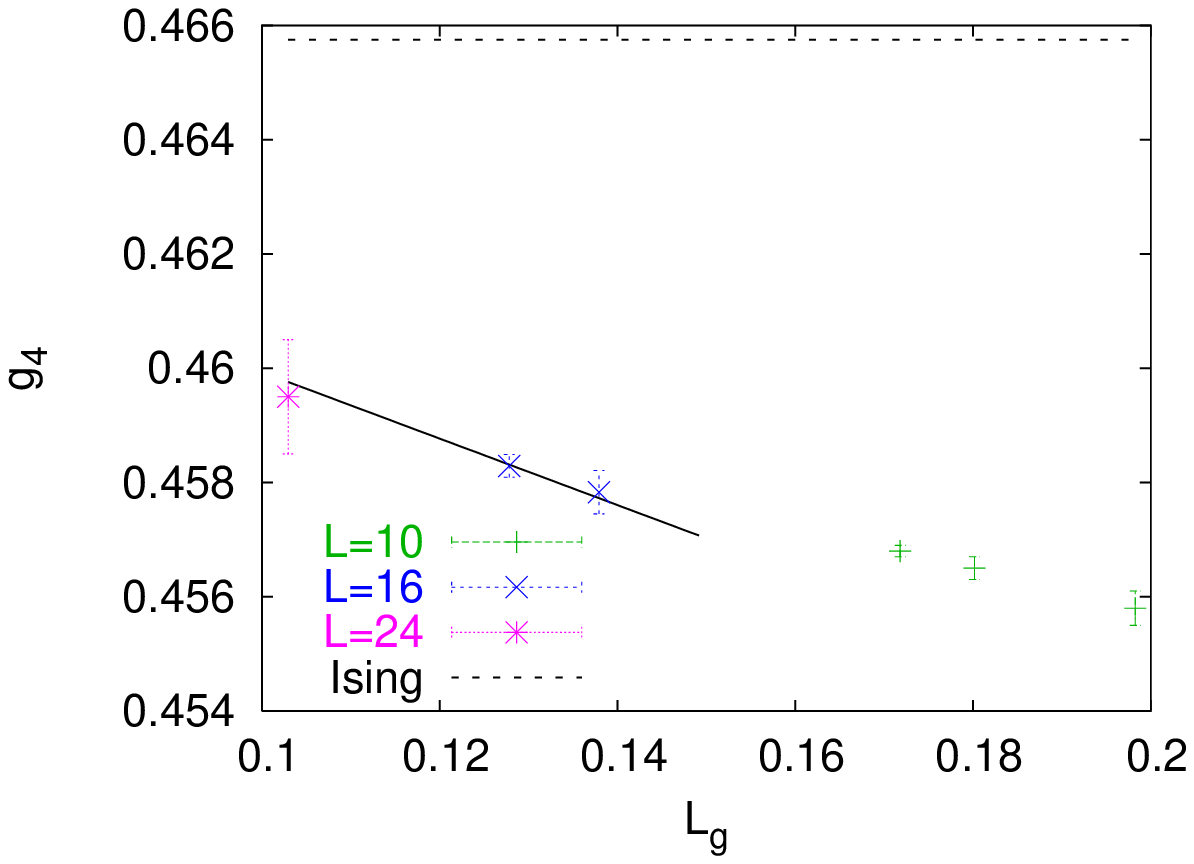}
\caption{Same as Fig. \ref{fig:nt4g4}, but for $N_\tau=2$ lattice.}
\label{fig:nt2g4}
\end{figure}
For the other intersection coordinate $g_4^*$ there is a noticeable scaling behavior, therefore we perform 
the FSS analysis analogous to the presented earlier ($N_\tau=4$). The results are plotted in Fig.
\ref{fig:nt2g4}. It is interesting that the smaller $N_\sigma=10$ lattice is not consistent with the scaling behavior, while it was for $N_\tau=4$ lattice. This can indicate that the sub-leading effects can be stronger here. For the larger lattices, however, the fit is good and indicates that the assumed $3D$ Ising universality class is correct. Also the assumption of the $3D$ Ising universality class is suported by independent studies of critical exponent $\nu$ \cite{Papa:2002gt}. 

It is expected that for lattices
with smaller $N_\tau$ it is sufficient to consider lattices with correspondingly smaller $N_\sigma$ values, therefore it is
surprising that the intersection point of $N_\sigma=32$ and 24 lattices is located statistically significantly  below the Ising $g^{Ising}_4$ value and one needs even larger lattices to reach the value corresponding to the thermodynamic limit.

For $N_\tau=2$ we found several estimates in the most recent literature all in good agreement with our result. On $2\times 12^3$ lattice the estimate is
$\beta_c=1.877$ (no error given, presumably $1.877(1)$) \cite{Gavai:1996nt}, while more recent estimate from  $2\times 30^3$ and $2\times 40^3$ lattices gives $\beta_c=1.8735(4)$ \cite{Fortunato:2000ge,Fortunato:2000hg}. 

\subsection{$N_\tau=1$ lattice}
Here we look
at even more exceptional $N_{\tau}=1$ lattice. This lattice is very special and allows for two formulations. One formulation is motivated by considering the sum over plaquettes in the partition function. It is natural to assume that there is one time-like plaquette for each space coordinate. This case
requires special treatment of space-like links since they have conjugate staples only in one time direction
(we choose positive direction) unlike two directions (positive and negative) for other directions.
The other formulation can be motivated by considering the $N_\tau=2$ lattice and performing factor
2 decimation in time direction (removing one time slice of links). In this formulation 
there is no special treatment of conjugate staples of space-like links. One has to consider staples in both positive and negative direction for each direction (including the  time direction).

We start with the first formulation of $N_\tau=1$ lattice, which we call time-like plaquette single counting.
Afterwards we will consider the second formulation which we refer to as time-like plaquette double counting.

\begin{figure}[ht]
\includegraphics[width=\columnwidth]{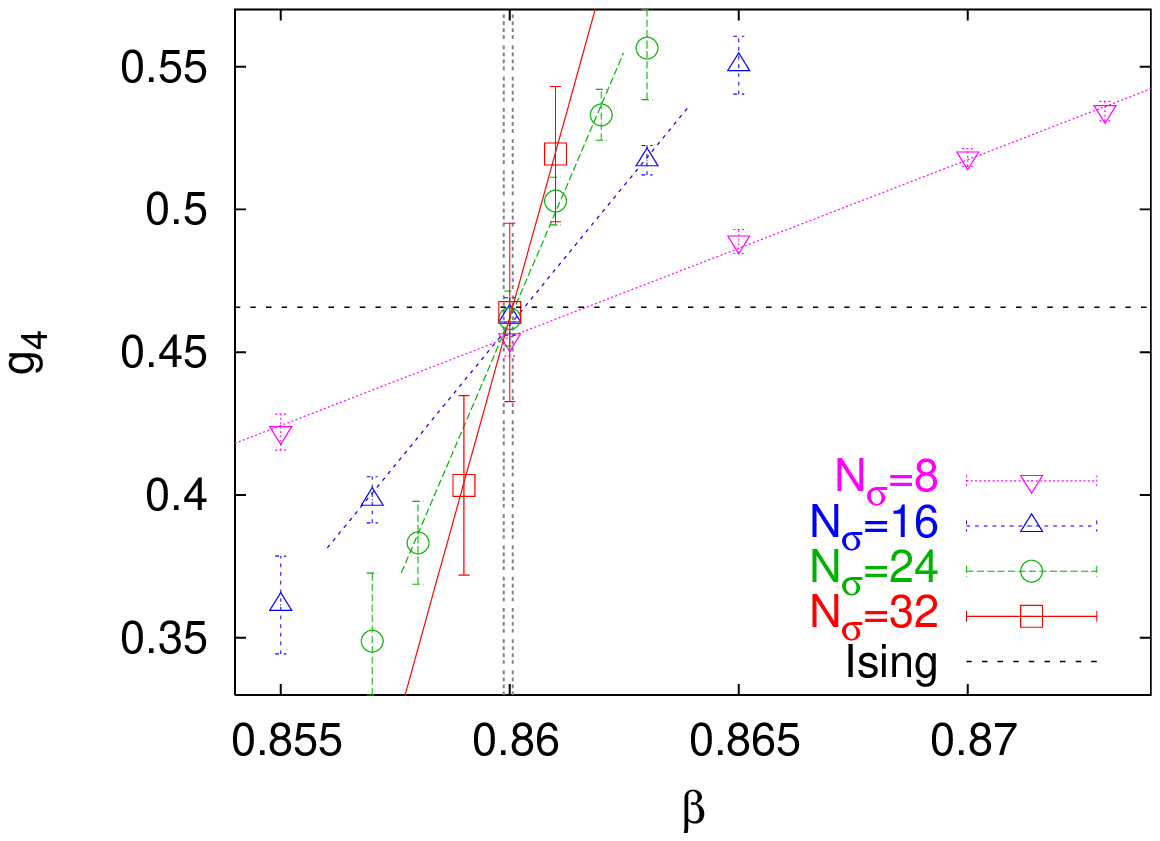}
\caption{The Binder cumulant $g_4$ for $N_\tau=1$ (single counting), $N_\sigma=8,16,24,32$ lattices. Linear fits are represented by solid lines, 
the $T_c$ estimate with errors by vertical lines.}
\label{fig:BinderC1s}
\end{figure}

For single counting formulation we present the results in Fig. \ref{fig:BinderC1s}.
Similarly to $N_\tau=2$ case we do not observe any scaling behavior since the lines cross at a single point. The linear fitting lines intersection coordinates are:
\begin{eqnarray}
 N_\sigma=8, N_\sigma^\prime=16 &-& \beta^*=0.85969(13), g^*_{4}=0.4535(20),\nonumber\\
 N_\sigma=16, N_\sigma^\prime=24 &-& \beta^*=0.85989(11), g^*_{4}=0.4573(28),\nonumber\\
 N_\sigma=24, N_\sigma^\prime=32 &-& \beta^*=0.85997(10), g^*_{4}=0.4606(51).\nonumber
 \end{eqnarray}

Note little change in the $\beta^*$ and $g^*_4$ coordinates, there is virtually no scaling to be extracted here. 
The intersection of the largest lattices (32 and 24) defines the critical coupling $\beta_c=0.85997(10)$.
The Binder cumulant is within one sigma from the Ising value.

\begin{figure}[ht]
\includegraphics[width=\columnwidth]{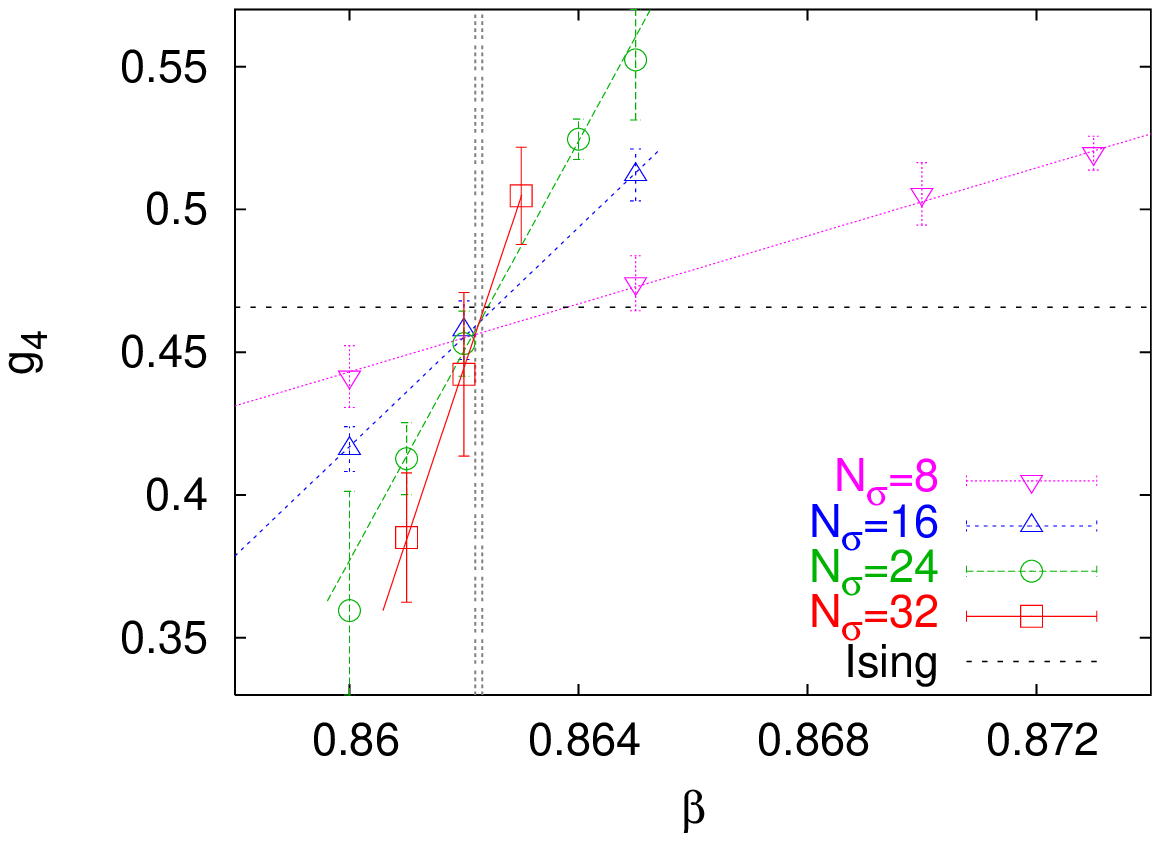}
\caption{The Binder cumulant $g_4$ for $N_\tau=1$ (double counting), $N_\sigma=8,16,24,32$ lattices. Linear fits are represented by solid lines, 
the $T_c$ estimate with errors by vertical lines.}
\label{fig:BinderC1d}
\end{figure}

Next we study the second formulation with time-like plaquettes double counting. We summarize the
results in Fig. \ref{fig:BinderC1d}. The pairwise fitting lines intersection coordinates in this case are:
\begin{eqnarray}
N_\sigma=8,  N_\sigma^\prime=16 &-& \beta^*=0.86198(11), g^*_{4}=0.4549(22),\nonumber\\
N_\sigma=16, N_\sigma^\prime=24 &-& \beta^*=0.86228(6), g^*_{4}=0.4606(12),\nonumber\\
N_\sigma=24, N_\sigma^\prime=32 &-& \beta^*=0.86226(6), g^*_{4}=0.4598(31).\nonumber
\end{eqnarray}
Again there is almost no scaling behavior in the coordinates.  The values of Binder cumulant are comparable to the previous formulation, however the inverse critical coupling is shifted toward larger values.
Also we observe that the Binder cumulant deviation from the Ising value is small ($<2\sigma$ away). 

The only known result in the recent literature for $N_\tau=1$ lattice is $\beta_c=0.8730(2)$ \cite{BenAv:1991ve}. However, the authors do not indicate how it was obtained. 

\subsection{$N_\tau=8$ lattice}
\begin{figure}[ht]
\includegraphics[width=\columnwidth]{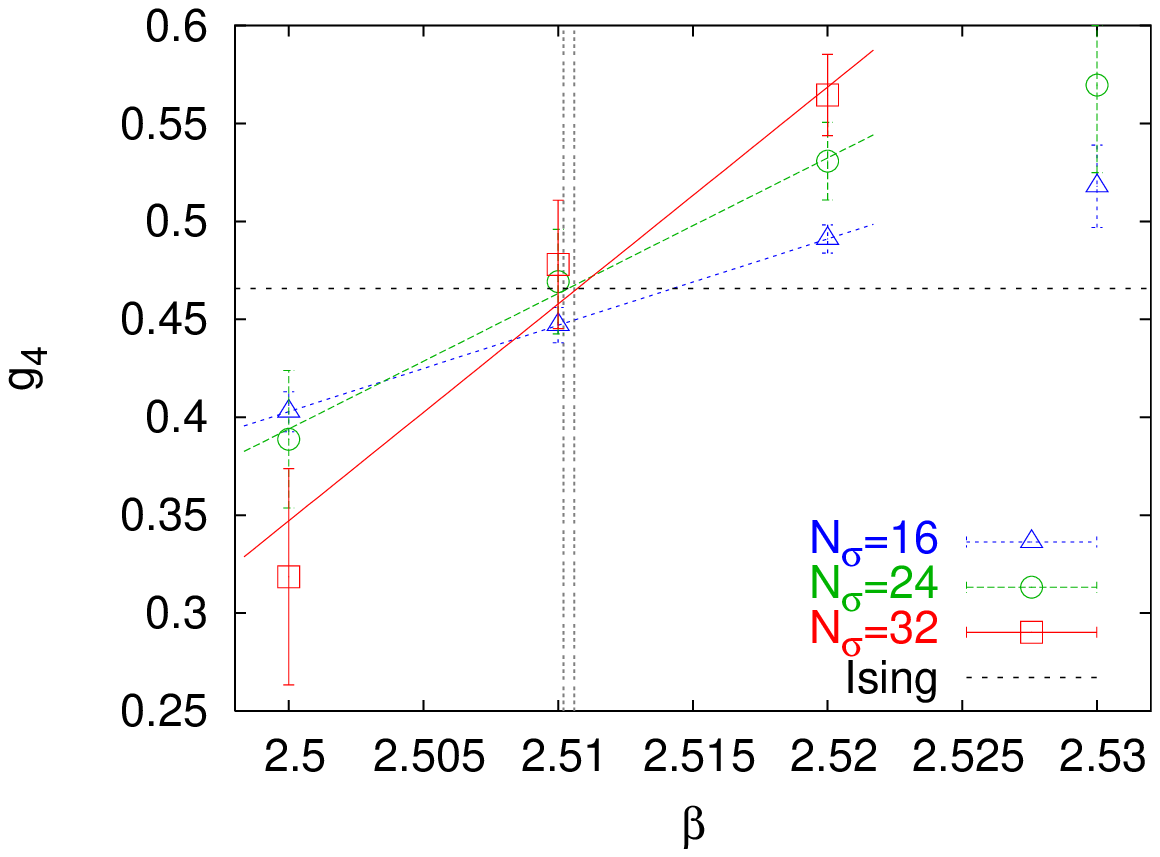}
\caption{The Binder cumulant $g_4$ for $N_\tau=8$, $N_\sigma=16,24,32$ lattices. 
Linear fits are represented by solid lines, the $T_c$ estimate with errors by vertical lines.}
\label{fig:BinderC}
\end{figure}

Next we look at $N_\tau=8$ and $N_\sigma=16,24$ and $32$ lattices. We present the results in Fig. \ref{fig:BinderC}. 
 For lattice $N_\sigma=16$ we performed $10^6$ sweeps, while for $N_\sigma=24$ and
$32$ lattices we performed $0.2\times 10^6$ sweeps
(measuring every 20) at the three beta values closest to the transition. 

The approach adopted for smaller lattices is not very efficient here. The fitting lines for larger two
lattices intersect at
\begin{eqnarray}
N_\sigma=24, N_\sigma^\prime=32 &-& \beta=2.5113(19), g_4=0.472(13).\nonumber
\end{eqnarray}
As one can see the errors are quite significant here and better statistics is needed.

On the other hand if one assumes the universallity class of the $3D$ Ising model it is possible to obtain better results. Our strategy is similar to one used in Ref. \cite{Bogolubsky:2004gi}.
We look at the points of intersection of linear fitting curves with the $3D$ Ising Binder cumulant value
\begin{eqnarray}
N_\sigma=16 &\beta=2.5143(0)\nonumber\\
N_\sigma=24 &\beta=2.5104(2)\nonumber\\
N_\sigma=32 &\beta=2.5107(8).\nonumber
\end{eqnarray}

The crossing of the fitting line with the Ising value
defines the transition point. Note that $N_\sigma=16$ lattice intersects with $N_\sigma=24$ slightly off, 
while $N_\sigma=24$ and $N_\sigma=32$ at the Ising value. This indicates that the thermodynamic limit has set in for $N_\sigma\ge24$.
The uncertainties of the fitting parameters contribute to the error of the critical coupling.
Possibly the best result is for $N_\sigma=24$ lattice since the larger lattice is a bit noisier, therefore we define $\beta_c=2.5104(2)$.

We compare our result to a similarly performed study of Ref. \cite{Bogolubsky:2004gi}, which is
$\beta_c=2.5105(10)$ for $N_\tau=8$ and $N_\sigma=32,40,48$ lattices.
The Gaussian difference test yields $Q=0.92$, indicating good correspondence of the results.

\section{String tension and physical scale\label{sec:string}}
In the previous section by measuring the critical Binder cumulant value we obtained high precision critical coupling values for various $N_\tau$ lattices. In this section we measure the Polyakov loop correlator and correspondingly obtain the static quark-antiquark potential. The results of the previous section make possible to study the scaling of dimensionless ratio $T_c/\sqrt{\sigma}$ well into the strong coupling regime ($\beta=1.87380$).

To fix the scale we use the critical coupling estimates of the previous section together with the value from the literature \cite{Bogolubsky:2004gi} for $N_\tau=12$ lattice. The critical temperature $T_c=1/(N^c_\tau*a)$ fixes the lattice spacing in physical units. 
The lattice is then simulated at these $\beta_c$ values but at different $N_\tau=3\cdot N^c_\tau$ corresponding to $T=T_c/3$. At this temperature finite temperature correction should be
minimal. 

For the string tension measurements we use L{\"u}scher-Weisz multilevel algorithm \cite{Luscher:2001up,Luscher:2002qv}.
As was noted in the second reference the one level algorithm is computationally
preferable. In our simulations the lattice
is sliced into layers of thickness $2a$. On each layer we perform 10 sweeps for $N_\tau/3=12$ lattice, 100 sweeps for  $N_\tau/3=8$ lattices, and 1000 sweeps for $N_\tau/3=4$ and 2 lattices.

We fit the potential using the following ansatz
\begin{equation}
\hat V(\hat r)=\hat V_0-\frac{\hat \mu}{\hat r}+\hat\sigma \hat r,\label{eq:VFansatz}
\end{equation}
where the hats indicate lattice dimensionless observables.

The fit of the potential is performed in the range $\hat r\in[2,5-9]$, which  ensures that no short distance $O(1/r^3)$ or long distance (reduced signal/noise or effects
from propagation across the periodic boundary condition) artifacts contribute.

\begin{table}[ht]
\centering
\begin{tabular}{cc|ccc|c}
$N_\tau/3$&$\beta$&$\hat V_0$&$\hat\mu$&$\hat\sigma$&$T_c/\sqrt{\sigma}$\\
\hline
12&2.6355(10)&0.524(28)&0.273(35)&0.0153(53)&0.67(12)\\
8&2.51098(58)&0.5505(10)&0.2757(12)&0.03232(18)&0.6953(19)\\
4&2.29850(6)&0.5826(6)&0.3241(7)&0.13312(10)&0.6852(03)\\
2&1.87380(3)&0.313(13)&0.256(15)&0.6285(26)&0.6307(13)\\
\end{tabular}
\caption{\label{tab:str_data} The parameters of the potential fit in lattice units and the 
string tension in (physical) units of $T_c$ for various $N_\tau/3\equiv N_\tau^c$ lattices.}
\end{table}
In table \ref{tab:str_data} we collect the data from the potential fits to the ansatz (\ref{eq:VFansatz}). 
The typical goodness of fit $Q\in [0.42,0.99]$.
Note that for $N_\tau/3=2,4,8$ lattices we performed the simulations at $\beta$ values from our earlier estimates, which are slightly different from the final values presented in the previous section. For the errors on the inverse coupling we take either the difference to the final value or the final value error, depending on which one is larger. 

In order to compare the string tension $\hat\sigma$ obtained on different lattices, we need to convert the lattice
observable to physical units. The string tension in physical units is $\sigma=\hat\sigma/a^2$. We can express
the lattice spacing in physical units through the critical temperature $a=1/(T_c\cdot N_\tau^c)$. Therefore we
can construct a dimensionless observable
$T_c/\sqrt{\sigma}=(N_\tau^c\cdot\sqrt{\hat\sigma})^{-1}$, which we present in the last column of the table.
The uncertainties come from the estimates of $a$ and  the string tension $\hat\sigma$
\begin{equation}
\frac{T_c}{\sqrt{\sigma}}=\frac{T_c a}{\sqrt{\hat\sigma}}.
\end{equation}
Therefore the error of this observable is 
\begin{equation}
\delta(\frac{T_c}{\sqrt{\sigma}})=\left(\left(\frac1{8 N_\tau^c b_0}\delta\beta_c\right)^2+
\left(\frac1{2N_\tau^c\hat\sigma^{3/2}}\delta\hat\sigma\right)^2\right)^{1/2}
\end{equation}
where $b_0=11N_c/(48\pi^2)=11N_c/(24\pi^2)$ and comes from the scaling of the lattice spacing with the coupling in the 
continuum limit\footnote{Here we consider only the leading term of (\ref{eq:a_scale}).}. We assume absence of correlations between measurements of critical coupling and string tension.

The data indicates that the string tension values in physical units are consistent for lattices $N_\tau/3\ge4$, although $N_\tau/3=12$ value obviously needs better statistics. The smaller $N_\tau/3=2$ value is significantly lower than other values. This is clearly related to the violation of the hyper-scaling of observables as the lattice coupling is increasing. Here we observe the change from the weak coupling to strong coupling regimes. The scaling window starts around $\beta\sim 2.29850$ ($N_\tau/3=4$).
It is interesting that the violation of scaling at $\beta=1.87380$ is relatively small ($<10\%$) for this ratio.
Note also that $V_0$ in physical units scales like $1/a$, therefore $\hat V_0$ is approximately constant $\sim0.5-0.6$
except for $N_\tau^c=2$ where it is significantly lower. 

\section{Summary}
We systematically studied $N_\tau=1,2,4$ and 8 finite temperature $SU(2)$ lattice gauge theory. 
The measurement of Polyakov loops in the vicinity of the transition point allowed us to study the scaling of the Binder cumulant.  We found that the critical values of Binder cumulant correspond to the $3D$ Ising model universality class. 

\begin{table}[ht]
\centering
\begin{tabular}{c|c}
$N_\tau$&$\beta_c$\\
\hline
16&2.7310(20)$^*$\\
12&2.6355(10)$^*$\\
8&2.5104(2)\\
6&2.4265(30)$^{\dagger}$\\
4&2.2991(2) $^{\prime}$, 2.2993(3)$^{\prime\prime}$\\
2&1.87348(2)\\
1(s)&0.85997(10) \\
1(d)&0.86226(6)\\
\end{tabular}
\caption{\label{tab:bc} Critical inverse coupling $\beta_c$ for 
different $N_\tau$ lattices: $^*$ is from \cite{Bogolubsky:2004gi}, $^{\dagger}$ is from \cite{Engels:1992fs},
$^{\prime}$ is our FSS estimate, while $^{\prime\prime}$ is obtained from the intersection of the largest lattices.}
\end{table}
New high precision estimates of the inverse critical coupling are obtained and summarized for various
$N_\tau$ lattices in Tab. \ref{tab:bc} together with estimates from the literature for lattices where we did not perform measurements. In particular we present two formulations for $N_\tau=1$ lattice with results different from the previous estimates.

From the study of static quark-antiquark correlators we extracted the string tension and using the critical couplings were able to obtain the dimensionless quantity $T_c/\sqrt{\sigma}$. This quantity shows small scaling violations for $N_\tau/3=2$ lattice ($N_\tau^c=2$, $\beta=1.87380$) and virtually no violations for larger $N_\tau$ lattices (weaker coupling).

\section*{Acknowledgments}
We thank Academic Technology Services (UCLA)
for computer support. The author would like to acknowledge insightful comments from Yannick Meurice and Peter Petreczky.
This work was supported by the Joint Theory Institute funded together by 
Argonne National Laboratory and the University of Chicago.
This work is supported in part by the U.S. Department of Energy, Division of High Energy Physics and Office of Nuclear Physics, under Contract DE-AC02-06CH11357.

\bibliographystyle{hunsrt}
\bibliography{avref}

\end{document}